\documentclass[a4paper,10pt,twoside, titlepage,openright,colorlinks]{article}
\def\figtype{2} 
\input epsf

\usepackage[usenames,dvipsnames]{color}
\usepackage[pdftex]{hyperref}
\definecolor{DarkRed}{RGB}{195,0,0} 
\hypersetup{colorlinks,%
            citecolor=OliveGreen,%
            filecolor=cyan,%
            linkcolor=DarkRed,%
            urlcolor=blue,%
            pdftex}
\usepackage[latin1]{inputenc}
\usepackage[english]{babel}
\ifnum\figtype=1 \usepackage[dvips]{graphicx} \fi
\ifnum\figtype=2 \usepackage{graphics} \fi
\usepackage{amsfonts,amssymb,amsmath,latexsym,epsfig}
\usepackage{verbatim}
\usepackage{natbib}
\usepackage{wasysym}
\usepackage{makeidx}
\usepackage{sidecap}
\usepackage{fancyhdr}
\usepackage[bf]{caption}
\usepackage{fix-cm}

\newcommand{\aj}[3]   {\href{http://adsabs.harvard.edu/cgi-bin/nph-abs_connect?version=1&warnings=YES&partial_bibcd=YES&sort=BIBCODE&db_key=ALL&bibstem=aj&year=#1&volume=#2&page=#3&nr_to_return=100&start_nr=1}{#1, AJ, #2, #3}}
\newcommand{\apjl}[3] {\href{http://adsabs.harvard.edu/cgi-bin/nph-abs_connect?version=1&warnings=YES&partial_bibcd=YES&sort=BIBCODE&db_key=ALL&bibstem=apjl&year=#1&volume=#2&page=L#3&nr_to_return=100&start_nr=1}{#1, ApJL, #2, L#3}}

\newcommand{\apj}[3]  {\href{http://adsabs.harvard.edu/cgi-bin/nph-abs_connect?version=1&warnings=YES&partial_bibcd=YES&sort=BIBCODE&db_key=ALL&bibstem=apj&year=#1&volume=#2&page=#3&nr_to_return=100&start_nr=1}{#1, ApJ, #2, #3}}

\newcommand{\arxiv}[2]{\href{http://arxiv.org/abs/#2}{#1 (arXiv:#2)}}

\newcommand{\mnras}[3]{\href{http://adsabs.harvard.edu/cgi-bin/nph-abs_connect?version=1&warnings=YES&partial_bibcd=YES&sort=BIBCODE&db_key=ALL&bibstem=mnras&year=#1&volume=#2&page=#3&nr_to_return=100&start_nr=1}{#1, MNRAS, #2, #3}}

\newcommand{\nhi}{n_{\textrm{{\scriptsize H}{\tiny \hspace{.1mm}I}}}}
\newcommand{\nhii}{n_{\textrm{{\scriptsize H}{\tiny \hspace{.1mm}II}}}}

\newcommand{\nd}{n_{\textrm{d}}}
\newcommand{\F}{F}
\newcommand{\Flam}{\F(\lambda)}
\newcommand{\Iem}{J_{\mathrm{em}}(\lambda)}
\newcommand{\Iobs}{J_{\mathrm{obs}}(\lambda)}
\newcommand{\T}{\mathcal{T}}

\renewcommand{\sec}[1]{Sec.~\ref{sec:#1}}

\newcommand{\fig}[1]{Fig.~\ref{fig:#1}}
\newcommand{\Fig}[1]{Figure \ref{fig:#1}}
\newcommand{\tab}[1]{Tab.~\ref{tab:#1}}
\newcommand{\Tab}[1]{Table \ref{tab:#1}}

\renewcommand{\cap}{\footnotesize}

\fancypagestyle{plain}{%
  \fancyhead{} 
 }

\begin{document}
\thispagestyle{empty}
\begin{center}
{\large Documentation for the intergalactic
        radiative transfer code}\vspace{3mm}\\
{\Huge {\sc IGMtransfer}}\vspace{2mm}\\
Version 1.1\vspace{7mm}\\
{\large Peter Laursen$^{1,2}$}\vspace{5mm}\\
$^1$ {\small Oskar Klein Centre, Dept.~of Astronomy, Stockholm University,
     AlbaNova, SE-106\,91 Stockholm, Sweden.}\\
$^2$ {\small Dark Cosmology Centre, Niels Bohr Institute, University of Copenhagen,
     Juliane Maries Vej 30, DK-2100, Copenhagen {\O}, Denmark,
\href{mailto:pela@dark-cosmology.dk?subject=Peter_Your_Code_Is_Really_Awesome_But}
{{\tt pela@dark-cosmology.dk}}.
}
\end{center}
\ \\
\ \\

\tableofcontents

\vspace{3.9mm}
{\bf Bibliography}
\vspace{-8mm}\begin{flushright}{\bf \pageref{bib}}\end{flushright}


\section{Introduction}
\label{sec:intro}

{\sc IGMtransfer} is a numerical code written in Fortran 90/95, intended for
simulating the radiative transfer (RT) of light in the vicinity of the
Ly$\alpha$ line through the intergalactic medium (IGM). Originally written with
the purpose of investigating
how the IGM close to Ly$\alpha$ emitting galaxies reshapes the emission line,
it can also be used to probe the general transmission properties of the IGM,
thus making it useful in simulations of cosmic reionization.

\begin{figure}[!t]
\centering
\ifnum\figtype=1 \includegraphics [width=0.90\textwidth] {r0.eps} \fi
\ifnum\figtype=2 \includegraphics [width=0.90\textwidth] {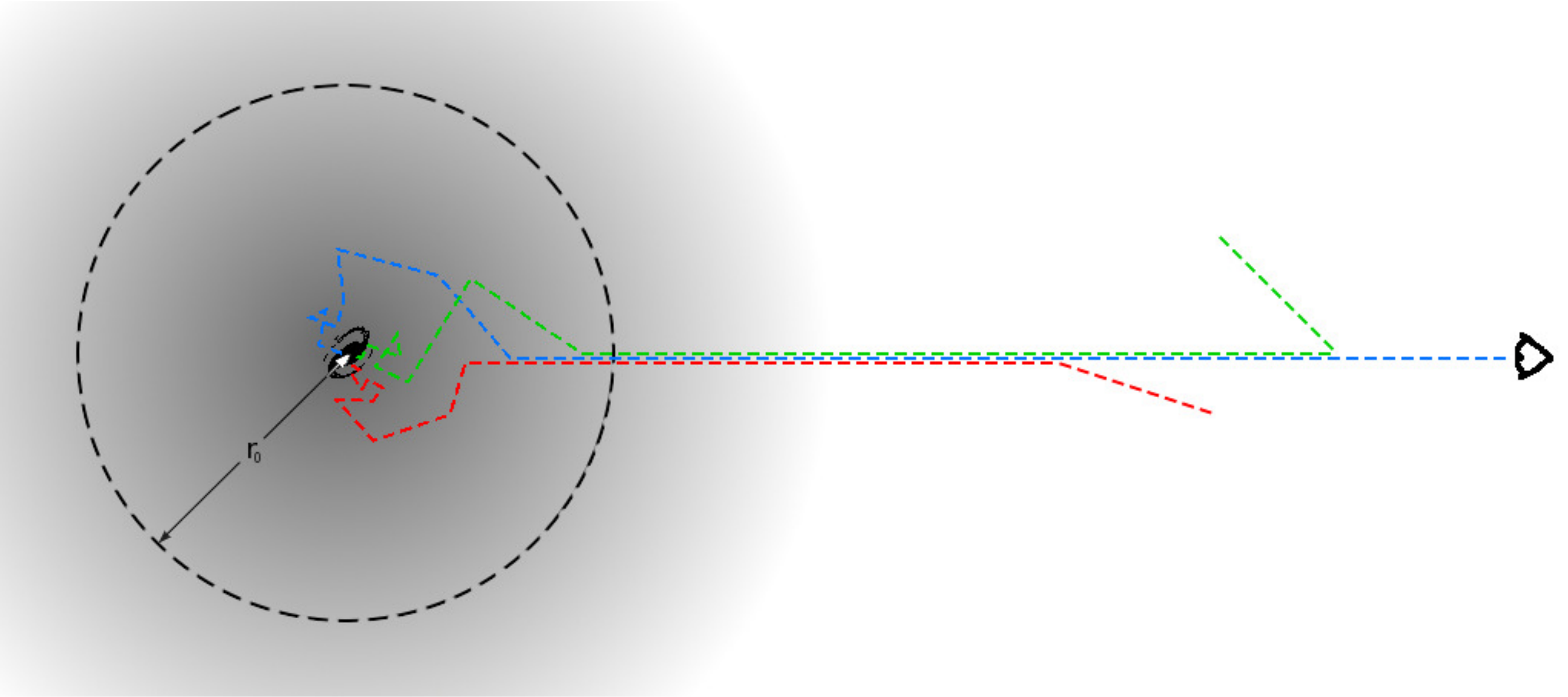} \fi
\caption{{\cap Illustration of the difference between the galactic RT and the
               IGM RT. Close to the galaxy, photons are scattered both in and
               out of the line of sight. In the rarefied IGM, photons are
               mainly scattered out of the line of sight, obviating the need
               for a full Ly$\alpha$ RT. The exact value of the distance $r_0$
               from a galaxy to begin the IGM RT is somewhat arbitrary, but is
               of the order of the virial radius of the galaxy.}}
\label{fig:r0}
\end{figure}

Ly$\alpha$ is a resonant line, meaning that a photon born in the gas surrounding
a hot star doesn't escape the galaxy until it has scattered millions of times
on neutral hydrogen, constantly changing direction and frequency.
To calculate the exact spectrum of Ly$\alpha$ photons escaping a galaxy, the
full resonant scattering RT needs to be solved. Once a Ly$\alpha$ photon is
the tenuous IGM, the probability of being scattered is much smaller, yet
in general not negligible.
Although the physics of scattering in galaxies and that of
scattering in the IGM is not inherently different, the difference in physical
conditions imposes a natural division of the two schemes: in the dense gas of
galaxies, photons are continuously scattered in and out of the line of sight,
whereas in the IGM, once a photon is scattered out of the line of sight, it
is ``lost'', becoming part of the background radiation. The probability of a
background photon being scattered \emph{into} the line of sight, on the other
hand, is vanishingly small. An illustration of this is seen in \fig{r0}.

If you haven't already done so, download all the necessary files from the URL
\href{http://www.dark-cosmology.dk/~pela/IGMtransfer.html}
    {www.dark-cosmology.dk/\~{}pela/IGMtransfer.html}.
Besides this documentation, the archive
\verb+IGMtransfer-+$[$\emph{version}$]$\verb+.tar.gz+ contains the
following files:

\begin{center}
\begin{tabular}{p{2.9cm}p{8cm}}
\verb+IGMtransfer.f90+     & Main code.\\
\verb+ProcessIGM.f90+      & Processes the output from {\sc IGMtransfer}.\\
\verb+F_lam.pro+           & IDL code visualizing the output from
                             {\sc ProcessIGM}.\\
\verb+fold[IPF].vim+       & Three Vim-scripts that structure the contents of
                             the above three codes, if your editor is Vim.\\
\verb+test.in+             & Example input file for {\sc IGMtransfer}.\\
\verb+testdir/+            & A subdirectory for example files containing the
                             following two files:\\
 \ -- \verb+CellData.bin+  & Example input data file for {\sc IGMtransfer},
                             containing the physical parameters of the gas in
                             a snapshot of a cosmological simulation at
                             $z = 3.5$.\\
 \ -- \verb+GalData.dat+   & Example input file containing physical parameters
                             of the galaxies in the snapshot.\\
\verb+toymodel.in+         & Input file for a small toy model.\\
\verb+toymodel/+           & A subdirectory containing the following three
                             files:\\
 \ -- \verb+CellData.dat+  & ASCII data file with toy model gas parameters.\\
 \ -- \verb+dat2bin.f90+   & Converts ASCII data to binary, ready for
                            {\sc IGMtransfer}.\\
 \ -- \verb+GalData.dat+   & Data for two toy galaxies.\\
\end{tabular}
\end{center}

After the following description of the basic principles and the physics of the
main code, the individual programs are explained.
A more thorough description is given in \citet{lau11}, which represents the
work first employing {\sc IGMtransfer}, and in \citet[][my Ph.D. thesis]{lau10}.

\subsection{Underlying concepts}
\label{sec:concept}

\begin{figure}[!t]
\centering
\ifnum\figtype=1 \includegraphics [width=0.50\textwidth] {bounce.eps} \fi
\ifnum\figtype=2 \includegraphics [width=0.50\textwidth] {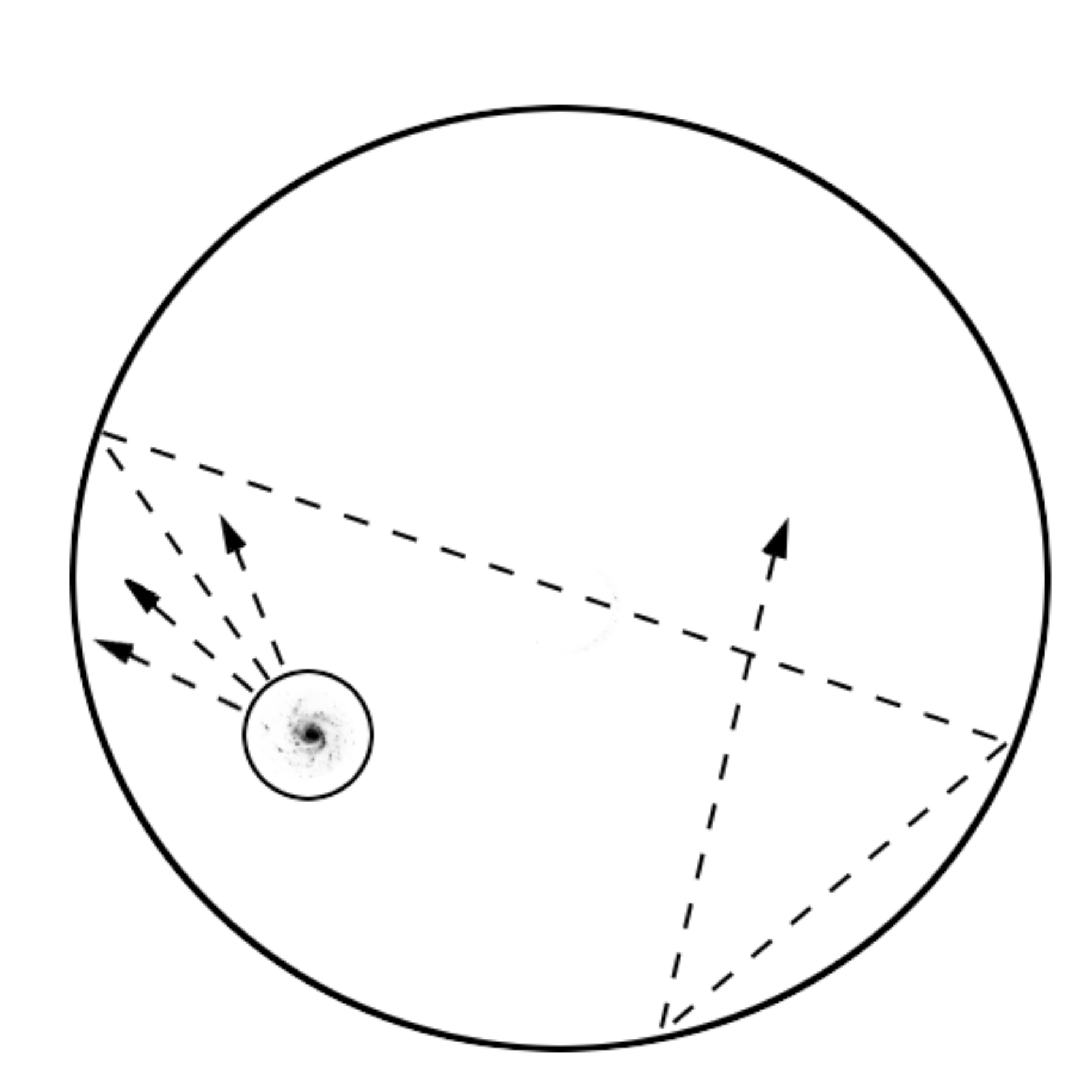} \fi
\caption{{\cap Illustration of how sightlines are cast through the cosmological
               volume (in the case of a spherical subvolume).
               To sample sufficiently the full solid angle of $4\pi$
               around the individual galaxies, {\tt n\_los} ($\sim$$10^3$)
               sightlines are cast from each galaxy (of which four are shown
               here). Each sightline is started at a distance $r_0$ --- which
               reflects the distance from the center at which the full
               scattering RT is no longer necessary --- from the
               center of a galaxy and traced until the bluest wavelength of the
               emitted spectrum has been redshifted into resonance. When the
               edge of the spherical volume is reached, the ray ``bounces''
               back, i.e.~continues in a random inward angle.}}
\label{fig:bounce}
\end{figure}

{\sc IGMtransfer} performs the IGM RT in a ``computational box'' with a
(possibly adaptively refined) cell-based structure.
The final results are obtained by considering the average of the RT performed
for many sightlines emerging in many
directions from many galaxies. For a given sightline, a (normalized) spectrum
is emitted, suffering random absorption lines (i.e.~the Ly$\alpha$ forest) as
it is continuously redshifted
when receding from the galaxy. When the edge of the computational volume
is reached, the sightline continues in a random, inward angle, thus ``bouncing''
around until the bluest wavelength of the simulated spectrum has been
redshifted into the Ly$\alpha$ resonance (\fig{bounce}).
{\sc IGMtransfer} was originally applied in a non-periodic, spherical volume.
The present version (v1.1) includes the possibility of using the full volume of
the computational box, and
a future version will include the possibility of utilizing periodic boundary
conditions, i.e.~once the edge is reached, the sightline continues on the
opposite side of the box.

The spectrum of each individual sightline is written to an output file which
can subsequently be processed by the program {\sc ProcessIGM}.
This two-step process also enables the user to use the calculated spectra to
perform other analyses, e.g.~investigate them for the relative abundance
of different absorption systems.


\subsection{Main output}
\label{sec:out}

The final, main output are the two related quantities:
\begin{enumerate}
\item the \emph{transmission function} $\Flam$, giving the fraction of light
      transmitted as a function of wavelength (as an average over many
      sightlines cast through the cosmological volume), and
\item the average transmission $\T$ of the IGM in a wavelength interval
      blueward of the Ly$\alpha$ line.
\end{enumerate}

\subsubsection{Transmission function}
\label{sec:Ftheo}

The resulting value of $\Flam$ at wavelength $\lambda$ for a given sightline is
\begin{equation}
\label{eq:F}
\Flam = e^{-\tau(\lambda)}.
\end{equation}
The optical depth $\tau$ is the sum of contributions from all the cells
encountered along the line of sight:
\begin{equation}
\label{eq:tau}
\tau(\lambda) = \sum_i^{\mathrm{cells}}
                r_i
                \,n_{\textrm{{\scriptsize H}{\tiny \hspace{.1mm}I}},i}
                \,\sigma(\lambda + \lambda v_{||,i}/c).
\end{equation}
Here,
$n_{\textrm{{\scriptsize H}{\tiny \hspace{.1mm}I}},i}$ is the density of
neutral hydrogen in the $i$'th cell,
$r_i$ is the distance covered in that particular cell,
$v_{||,i}$ is the velocity component of the cell along the line of sight,
and
$\sigma(\lambda)$ is the cross section of neutral hydrogen.
Due to the resonant nature of the transition, the largest contribution at a
given wavelength will arise from the cells the velocity of which corresponds
to shifting the wavelength close to resonance. However, for sufficiently
high column density absorbers (the so-called ``damped Ly$\alpha$ absorbers) the
damping wing of the profile may cause absorption at velocities quite far from
this.

If dust is present in the simulation, $\nhi\sigma(\lambda)$ is replaced by
$\nhi\sigma(\lambda) + \nd\sigma_{\mathrm{d}}(\lambda)$.

Assuming a spectrum of light $\Iem$ escaping a galaxy (simulated, e.g., using
Monte Carlo simulations, as in \citet{lau09a,lau09b}), the final, observed
spectrum $\Iobs$ is then
\begin{equation}
\label{eq:Iobs}
\Iobs = \Flam \, \times \, \Iem.
\end{equation}

\begin{figure}[!t]
\centering
\ifnum\figtype=1 \includegraphics [width=1.00\textwidth] {spXF.eps} \fi
\ifnum\figtype=2 \includegraphics [width=1.00\textwidth] {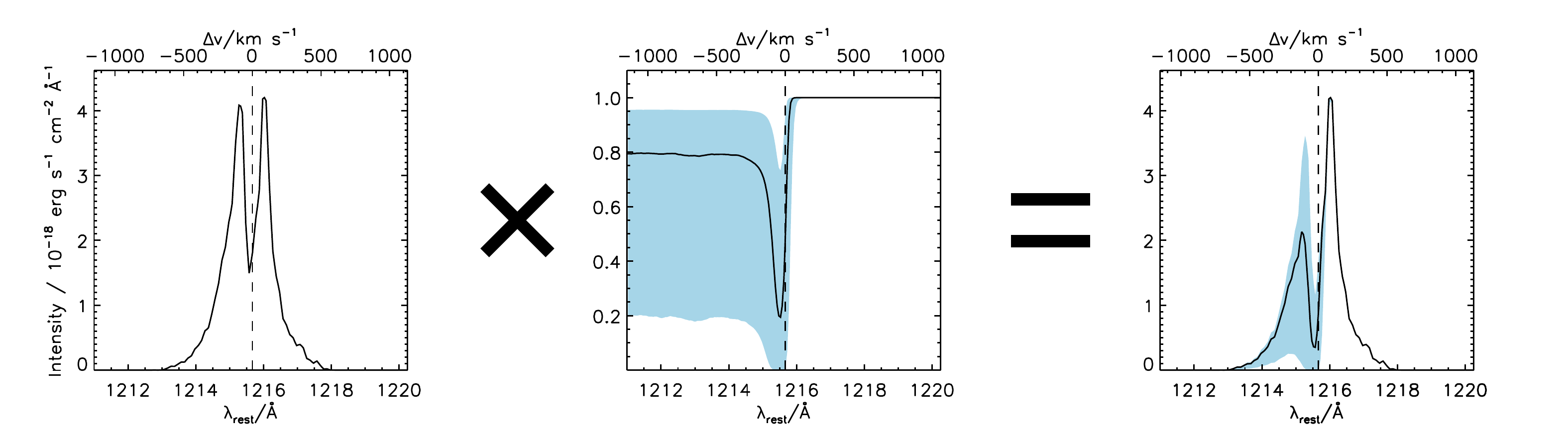} \fi
\caption{{\cap Illustration of the effect of the IGM on the observed Ly$\alpha$
         profile emerging from a galaxy at $z \sim 3.5$. Without taking into
         account the IGM, the two peaks are roughly equally high
         (\emph{left panel}). However, when the spectrum is transmitted through
         the IGM characterized by the transmission function $\Flam$
         (\emph{middle panel}, with the \emph{cyangrayish} region representing
         the 68\%
         confidence interval), the blue peak is dimished, resulting in an
         observed spectrum with a higher red peak (\emph{right panel}).
         The figure is taken from \citet{lau11}}.}
\label{fig:spXF}
\end{figure}

\Fig{spXF} illustrates this effect.
Due to the correlation of the IGM with the source, $\Flam$ is non-trivial close
to the Ly$\alpha$ line, but far from the source, or, spectrally speaking, at
very blue wavelengths, it becomes a constant function of wavelength (as long as
one does not enter an appreciably different redshift epoch).


\subsubsection{Average transmission}
\label{sec:Ttheo}

\begin{figure}[!t]
\centering
\ifnum\figtype=1 \includegraphics [width=0.80\textwidth] {Songaila.eps} \fi
\ifnum\figtype=2 \includegraphics [width=0.80\textwidth] {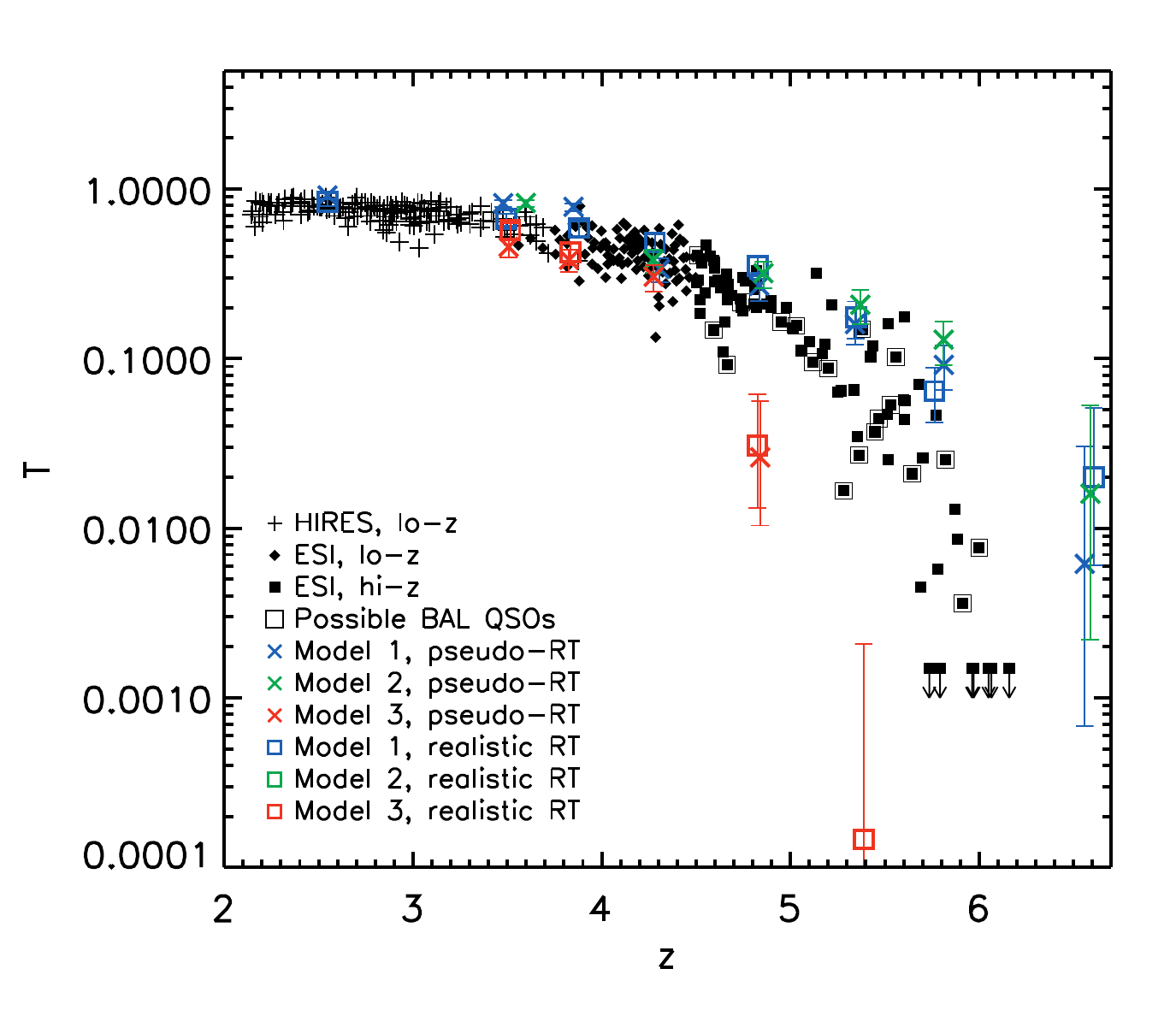} \fi
\caption{{\cap Comparison of observations (\emph{black data points}) and
               simulations (\emph{colored data points}) of the transmitted flux
               blueward of the Ly$\alpha$ line as a function of redshift.
               The figure is taken from \citet{lau11}, but the observed data
               points are from \citet{son04}.}}
\label{fig:Songaila}
\end{figure}

The transmission function is probably only accurate if you have very high
resolution in your simulation; if your cell size is of the order of the virial
radius of your galaxies, you will probably overestimate the absorption.
However, {\sc IGMtransfer} may still sweeten your life by calculating the
\emph{average}, or rather median, transmitted fraction $\T$ in a large
wavelength interval bluward of the Ly$\alpha$ line. \Fig{Songaila} shows such
calculated fractions at different redshifts, compared to the observations of
\citet{son04}.
%



\subsection{{\tt fold[IPF].vim}}
\label{sec:fold}

{\sc IGMtransfer} is written in one single file. If your editor is Vim,
\emph{first} time you open the source code use
\begin{verbatim}
   vim -s foldI.vim IGMtransfer.f90
\end{verbatim}
This will fold distinct parts of the code into single lines, making it more
manageable. To inspect the lines in a fold, go to that fold and press space.
To close the fold again,
press \verb+zc+ while standing in the fold. If you screw something up, exit and
open one more time with \verb+-s foldI.vim+.

Similarly, \verb+ProcessIGM.f90+ and \verb+F_lam.pro+ can be folded with
\verb+foldP.vim+ and \verb+foldF.vim+, respectively.

If the folds are gone when you open the files next time, try adding these
lines in your \verb+.vimrc+:
\begin{verbatim}
   au BufWinLeave * mkview
   au BufWinEnter * silent loadview
\end{verbatim}



\section{{\sc IGMtransfer}}
\label{sec:IGMtransfer}

\subsection{Input parameters}
\label{sec:in}

The main input parameters are given in an input file; for the example
simulation, this file is called \verb+test.in+. The individual input parameters
are explained below and summarized in \Tab{params}.
Additionally, two data files containing the physical parameters of the gas and
the galaxies, respectively, are required. The contents of these are explained
in \sec{celldata} and \sec{galdata}, respectively.\vspace{2mm}\\

\verb+indir+, \verb+subdir+, \verb+CellData+, and \verb+GalData+\\
The name of the files containing the gas data and the galaxy data are given in
the keyword \verb+CellData+ and \verb+GalData+, respectively. These files are
supposed to be put in a directory, the name of which is given by the keyword
\verb+subdir+ which, in turn, is a subdirectory of a mother directory given by
\verb+indir+.\vspace{1mm}\\

\verb+outdir+ and \verb+Ioutfile+\\
The output of {\sc IGMtransfer} is a file with the name given by the keyword
\verb+Ioutfile+. It is put in a directory, the name of which is also given by
\verb+subdir+, but which is a
subdirectory of the directory given by \verb+outdir+. If you want the data
files and the output to be in the same directory, simply give the same name to
\verb+indir+ and \verb+outdir+. Using two different directories, however, may
be useful if you wish to save the output on a desk which is backed
up.\vspace{1mm}\\

\verb+Poutfile+\\
The output from {\sc IGMtransfer} is processed by the program {\sc ProcessIGM}.
While the average transmission $\T$ (\sec{Ttheo}) is written to standard output,
the transmission function $\Flam$ (\sec{Ftheo}) is written in a text file, the
name which is given by the keyword \verb+Poutfile+. This data can be visualized
by means of the IDL code {\sc F\_lam}.\vspace{1mm}\\

\verb+n_write+\\
The number of sightlines traced between each time {\sc IGMtransfer} writes its
output.\vspace{1mm}\\

\verb+n_los+\\
The number of sightlines traced per galaxy.\vspace{1mm}\\

\verb+SpecRes+\\
Number of bins into which the spectral range is divided.\vspace{1mm}\\

\verb+BW+\\
Two-element vector giving the lower and upper values of the wavelength
interval in {\AA}ngstr\"om to be propagated through the IGM.\vspace{1mm}\\

\verb+BW_stat+\\
Two-element vector giving the lower and upper values of the wavelength
interval in {\AA}ngstr\"om to be used for calculating $\T$. In principle, the
lower value, \verb+BW_stat(1)+, could be set equal to \verb+BW(1)+.
However, due to the rather broad damping wing of Ly$\alpha$, as well
as the peculiar motion of the gas elements, absorption at a given wavelength is
caused by gas occupying a quite large range in real space. For this reason,
even though the spectrum is traced until \verb+BW(1)+ has been redshifted into
resonance, the average absorption will begin to decrease some {\AA}ngstr\"om
before. Thus, it is probably best to omit the bluest ten {\AA}ngstr\"om or so,
depending on the redshift\footnote{In real observations, the reason for not
using too blue wavelengths is both not to enter the Ly$\beta$ resonance and not
to enter an appreciably different redshift epoch; these complications are of
course not present in {\sc IGMtransfer}.}.

The higher value of \verb+BW_stat+ should be sufficiently far from the
Ly$\alpha$ line center that the state of the IGM --- i.e.~its density,
temperature, ionization state, and velocity field --- is not correlated with
the source. If you want to be, say, 1 Mpc away from your sources at
a redshift of 5, then $\Delta\lambda = \lambda_0 H(z=5)$ (1 Mpc) $/ c = 2.3$
{\AA}, i.e.~\verb+BW_stat(2)+ $\simeq 1213$ {\AA}.

If you want to compare to the sample of transmissions by \citet{son04},
use the interval $[1080,1185]$ {\AA}.
\vspace{1mm}\\

\verb+f_rvir+\\
As discussed in the introduction, the sightlines should be started sufficiently
far from the center of the galaxies that scattering \emph{into} the line of
sight is much less probable than scattering \emph{out of} the line of sight.
Denoting this distance $r_0$, if $r_0$ is too small, or too large, absorption
will tend to be over- or underestimated, depending on whether the immediate
surroundings of the galaxies on average cause more or less absorption than the
general IGM.
Of course $r_0$ depends on the galaxy in question, in particular its size, as
well as the general state of the circumgalactic gas. However, as argued in
\citet{lau11}, when measuring $r_0$ in terms of the virial radii of the
galaxies, $r_0$ for the different galaxies become comparable to each other,
and through a convergence study it was found that, at least at the redshifts
probed in that study ($z \sim 2.5$ to $\sim$6.5),
setting $r_0 = 1.5 r_{\mathrm{vir}}$ gave meaningful results.
\emph{The keyword} \verb+f_rvir+ \emph{specifies the number of virial radii at
which to start the sightlines}.\vspace{1mm}\\

\verb+r_eff+\\
{\sc IGMtransfer} was originally constructed to perform the RT in a spherical
subvolume of a box, with the radius of the sphere given by
\verb+r_eff+ $\in [0,1]$, the fraction of the half the side length of the box
(i.e.~the largest possible sphere has \verb+r_eff+ = 1). In order to use the
full box, simply set \verb+r_eff+ equal to any value larger than $\sqrt{3}$,
so that the sphere encloses the box fully.\vspace{1mm}\\

\verb+DustType+, \verb+f_ion+\\
If dust is included in the simulations, the string \verb+DustType+ should be
set to either `\verb+SMC+' or `\verb+LMC+', depending on whether the dust is
modeled as dust in the Small or Large Magellanic Cloud (SMC and LMC),
respectively. Motivated by observations, the dust density is
modeled as being proportional to the density of neutral hydrogen, plus a
fraction of the ionized hydrogen (as well as to the metallicity). This fraction
is given by the keyword \verb+f_ion+. See \sec{nd} on how to incorporate dust
in the simulation.\vspace{1mm}\\

\verb+z+, \verb+H_0+, \verb+Omega_M+, and \verb+Omega_L+\\
The redshift of the simulation, and the present values of the Hubble constant
(in km s$^{-1}$ Mpc$^{-1}$), matter, and cosmological constant parameters.
These values are used to convert the wavelength interval into a physical
distance.

\Tab{params} summarizes the input parameters.

\begin{table}[!h]
\begin{center}
{\sc Input parameters}\vspace{1mm}
\begin{tabular}{p{1.7cm}p{8.3cm}}
\hline
\hline
Parameter & Explanation \\
\hline
\verb+indir+$^\mathrm{c}$   & Mother directory for input data subdirectory.\\
\verb+subdir+$^\mathrm{c}$  & Subdirectory containing the data files.\\
\verb+CellData+$^\mathrm{c}$& Binary file containing physical parameters of the gas.\\
\verb+GalData+$^\mathrm{c}$ & Text file containing the galaxy data.\\
\verb+outdir+$^\mathrm{c}$  & Mother directory for output subdirectory.\\
\verb+Ioutfile+$^\mathrm{c}$& Output file of {\sc IGMtransfer}, and input file
                              for {\sc ProcessIGM}.\\
\verb+Poutfile+$^\mathrm{c}$& Output file of {\sc ProcessIGM}.\\
\verb+n_write+$^\mathrm{i}$ & Number of sightlines traced between each output.\\
\verb+n_los+$^\mathrm{i}$   & Number of sightlines per galaxy.\\
\verb+SpecRes+$^\mathrm{i}$ & Spectral resolution in \#bins.\\
\verb+BW+$^\mathrm{d}$      & Two-element vector giving the lower and upper
                              limit of the wavelength interval for the RT.\\
\verb+BW_stat+$^\mathrm{d}$ & Two-element vector giving the lower and upper
                              limit of the wavelength interval in which to
                              calculate $\T$.\\
\verb+f_rvir+$^\mathrm{d}$  & Number of virial radii from the center of a
                              galaxy, from where to begin a given sightline.\\
\verb+r_eff+$^\mathrm{d}$   & Radius of sphere in which to perform the RT, in
                              terms of half the side length of the box,
                              i.e.~$\in [0,1]$ for a spherical subvolume, or
                              $>\sqrt{3}$ to use the full box.\\
\verb+DustType+$^\mathrm{c}$& \verb+'SMC'+ or \verb+'LMC'+, depending on the
                              dust type.\\
\verb+f_ion+$^\mathrm{d}$   & Fraction of ionized hydrogen that contributes to
                              the dust density (see \sec{nd}).\\
\verb+z+$^\mathrm{d}$       & Redshift of snapshot.\\
\verb+H_0+$^\mathrm{d}$     & Hubble constant in km s$^{-1}$ Mpc$^{-1}$.\\
\verb+Omega_M+$^\mathrm{d}$ & Matter density parameter $\Omega_M$.\\
\verb+Omega_L+$^\mathrm{d}$ & Cosmological constant density parameter
                              $\Omega_\Lambda$.\\
\hline
\end{tabular}
\caption{{\small Explanation of parameters given in the input file
                {\tt test.in}. Superscripts $^\mathrm{c}$, $^\mathrm{d}$, and
                $^\mathrm{i}$ indicate character, real (double), and integer,
                respectively.}}
\label{tab:params}
\end{center}
\end{table}
%


\subsection{Gas data}
\label{sec:celldata}

The physical parameters of the intergalactic gas elements are given in
one-dimensional arrays in binary form in the file
\verb+indir+/\verb+subdir+/\verb+CellData+, in a nested, space-filling
hierarchy. Each array constitutes one record; these records are displayed in
\tab{CellData}, along with the first record that specifies a few important
numbers. The subsequent sections explain in further detail what the numbers
mean.
%
\begin{table}[!h]
\begin{center}
{\sc Contents of ``}{\tt CellData}''\vspace{1mm}
\begin{tabular}{p{4.7cm}p{6.5cm}}
\hline
\hline
Record                          & Explanation \\
\hline
\verb+N_cells+$^\mathrm{i}$,
\verb+D_box+$^\mathrm{d}$,
\verb+ni+$^\mathrm{i}$,
\verb+nj+$^\mathrm{i}$,
\verb+nk+$^\mathrm{i}$          & Total number of cells,
                                  total length of computational box in kpc,
                                  base grid resolution in $x$-, $y$-, and
                                  $z$-direction.\\
\verb+LevelString+$^\mathrm{i}$ & Refinement levels of cells. Optional.\\
\verb+n_HIString+$^\mathrm{s}$  & Neutral hydrogen number density per cm$^3$.\\
\verb+TString+$^\mathrm{s}$     & Temperature in Kelvin.\\
\verb+V_xString+$^\mathrm{s}$   & Gas bulk velocity in $x$-direction in km
                                  s$^{-1}$, \emph{in physical coordinates and
                                  with respect to the center of the box}.\\
\verb+V_yString+$^\mathrm{s}$   & Gas bulk velocity in $y$-direction.\\
\verb+V_zString+$^\mathrm{s}$   & Gas bulk velocity in $z$-direction.\\
\verb+ZString+$^\mathrm{s}$     & Metallicity in terms of Solar. Optional.\\
\verb+n_HIIString+$^\mathrm{s}$ & Ionized hydrogen number density per cm$^3$.
                                  Optional.\\

\hline
\end{tabular}
\caption{{\small Superscripts $^\mathrm{i}$, $^\mathrm{d}$, and $^\mathrm{s}$
                 indicate integer, double, and single precision real,
                 respectively.
                 All arrays ({\tt *String}) are of size {\tt N\_cells}.}}
\label{tab:CellData}
\end{center}
\end{table}

\subsubsection{{\tt N\_cells}, {\tt D\_box}, {\tt ni}, {\tt nj}, and {\tt nk}}
\label{sec:rec1}

The first record consists of the following five numbers:\\
\verb+N_cells+ is the total numbers of cells (of all refinement levels).\\
\verb+D_box+ is the length of the side of the computational box in \emph{proper
kpc}. That is, if your simulation is a box of, say, $10 h^{-1}$ comoving Mpc,
then for $h = 0.7$ a snaphot at $z = 3.46$ will have
\verb+D_box+ $= (10^3$ kpc)$\,/\,0.7\,/\,(1 + 3.46) = 3203.075$ kpc\footnote{The
example
simulation described in \sec{ex} is a spherical simulation of $10h^{-1}$
comoving Mpc at $z = 3.46$, but is actually enclosed in a slightly larger box
of {\tt D\_box} = 3400 kpc.}.\\
\verb+ni+, \verb+nj+, and \verb+nk+ is the spatial resolution of the base grid.
For a regular, unrefined grid,
\verb+N_cells+ = \verb+ni+$\times$\verb+nj+$\times$\verb+nk+.
 

\subsubsection{{\tt LevelString}}
\label{sec:level}

As already mentioned, {\sc IGMtransfer} includes the possibility of performing
the RT with ``adaptive mesh refinement'' (AMR);
that is, a grid where a number of cells
are split up into eight cells which, in turn, may be further split up into eight
cells, and so on, recursively, for an arbitrary number of times. The criterion
for splitting a cell will usually be a density threshold, but can in principle
be anything, e.g.~density gradient.

\begin{figure}[!t]
\centering
\ifnum\figtype=1 \includegraphics [width=0.70\textwidth] {AMR_2D.eps} \fi
\ifnum\figtype=2 \includegraphics [width=0.70\textwidth] {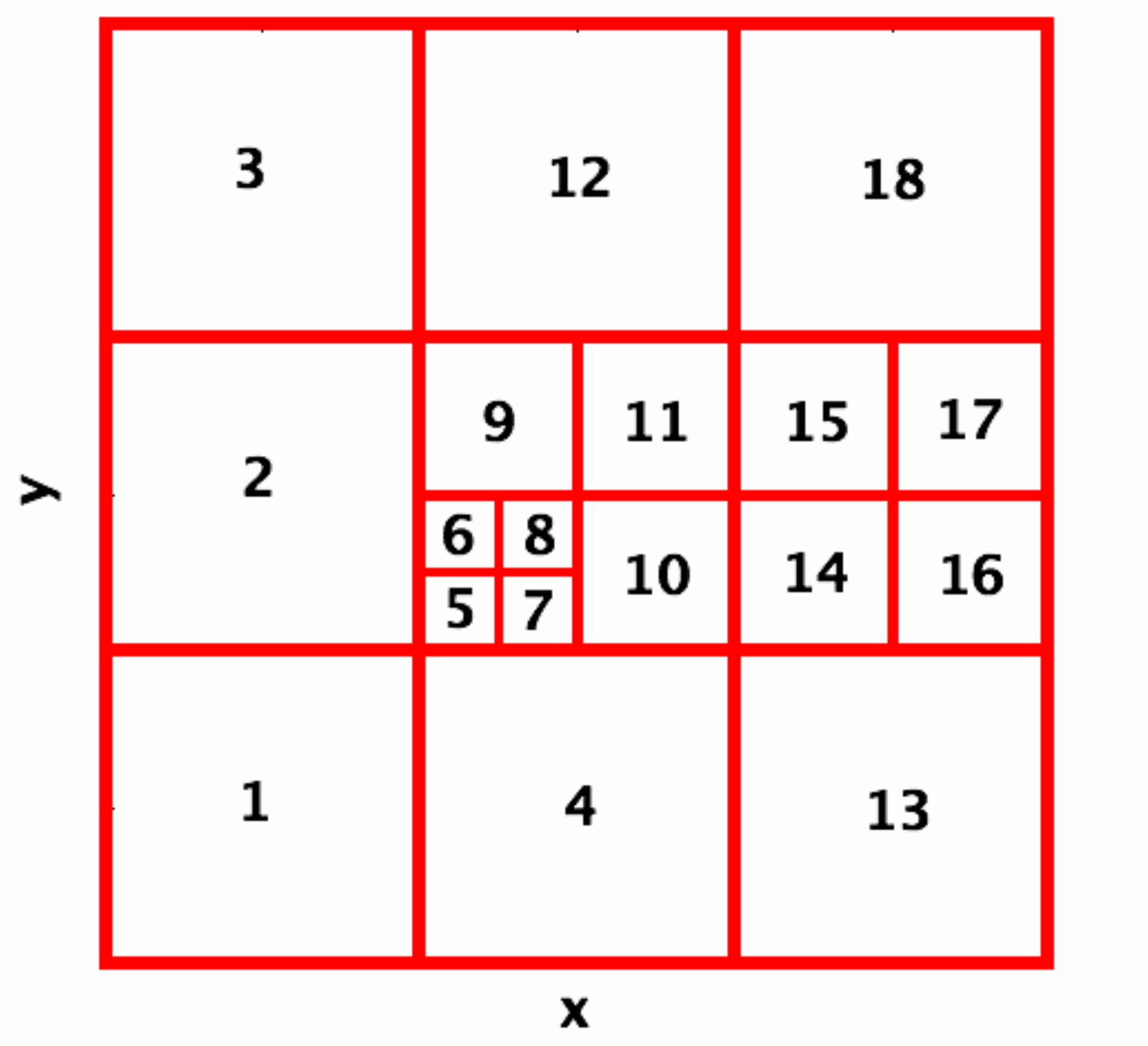} \fi
\caption{{\cap Two-dimensional example of how the AMR grid is represented in
               the code, with base grid resolution {\tt ni} = {\tt nj} = 3, and
               maximum refinement level 3.
               The space-filling curve goes through the grid first in
               the $y$-direction, and then in the $x$-direction. The order of
               cells are given by the indicated numbers. In the
               illustrated case, the array {\tt LevelString} would consist of
               the numbers\\
               $[$0 0 0 0 2 2 2 2 1 1 1 0 0 1 1 1 1 0$]$.}}
\label{fig:AMR_2D}
\end{figure}

\begin{figure}[!t]
\centering
\ifnum\figtype=1 \includegraphics [width=0.99\textwidth] {AMR_3D.eps} \fi
\ifnum\figtype=2 \includegraphics [width=0.99\textwidth] {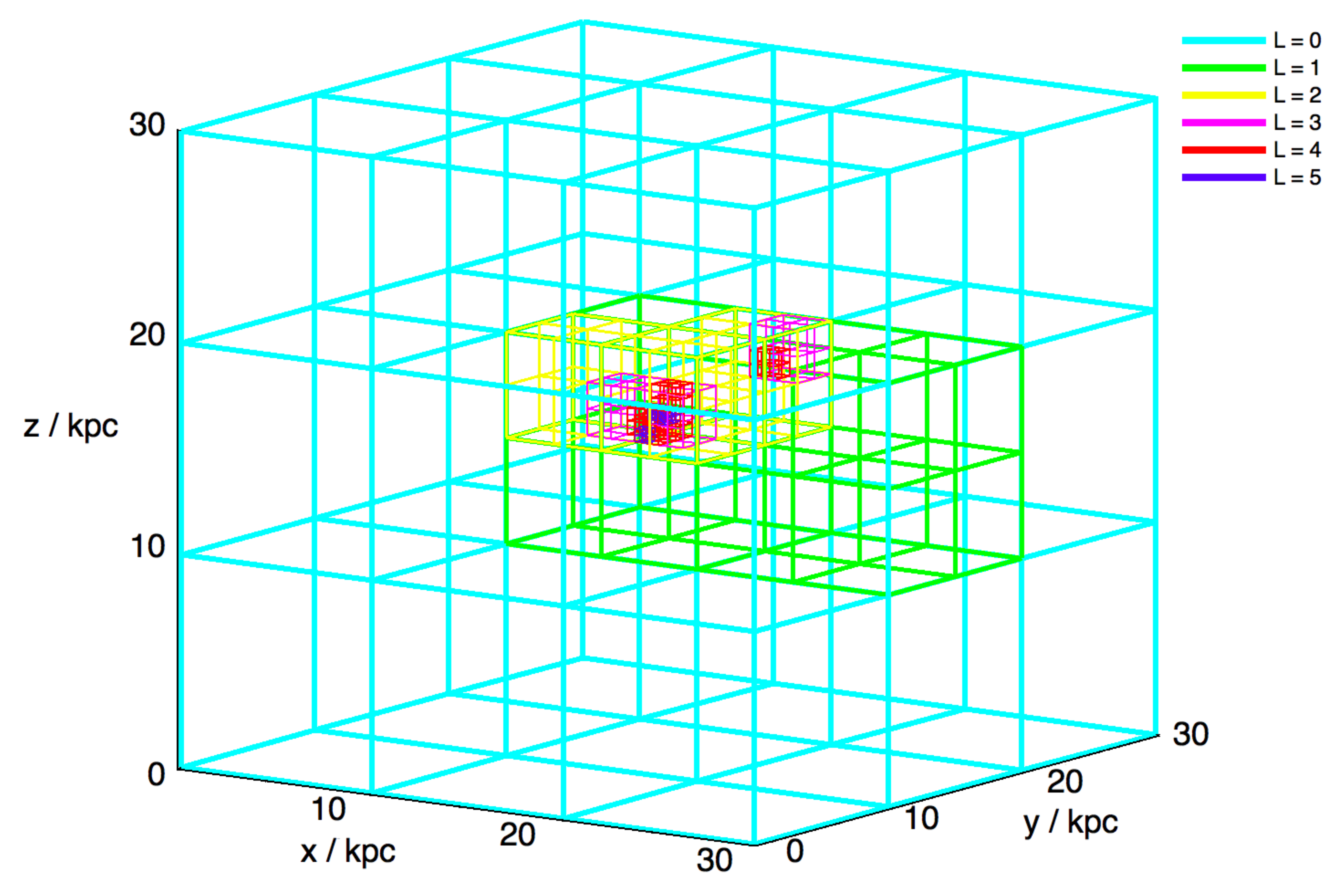} \fi
\caption{{\cap Three-dimensional toy model of a grid with five levels of
               refinement. Taking a look in the files
               {\tt toymodel.dat} and {\tt dat2bin.f90} should help you figure
               out the way the data is organized. See also
               \href{http://www.dark-cosmology.dk/~pela/Phd/AMRmovie.gif}
                    {www.dark-cosmology.dk/\~{}pela/Phd/AMRmovie.gif} for
               a 3D rotation.}}
\label{fig:AMR_3D}
\end{figure}

In the case of an adaptively refined grid, an array containing the
\emph{refinement level} of each cell must also be given.
The refinement of a cell is denoted by its refinement level $\mathcal{L}$,
where $\mathcal{L} = 0$ corresponds to the unrefined base grid.
Thus, if $dx_0$ is the length of a base cell, a cell
refined $\ell$ times has length $dx_0 / 2^\ell$. Proceeding first in the
$z$-direction, then in the $y$-direction, and finally in the $x$-direction,
the one-dimensional array \verb+LevelString+ contains the refinement levels of
each cell. If a cell is refined, this mapping continues in a recursive fashion
one level below.
\Fig{AMR_2D} illustrates the mapping in 2D,
while a 3D rendering of a toy
grid in seen in \fig{AMR_3D}. In the directory \verb+toymodel+, an ASCII file
\verb+toymodel.dat+ containing the relevant cell data for this toy grid is
found, along with a snippet \verb+dat2bin.f90+ that converts the data to
binary form, ready to be read in by {\sc IGMtransfer}.
This structure results in a unique ordering of the cells, and thus no $x$-,
$y$-, and $z$-positions are necessary.

The other physical parameters are given subsequently in separate arrays, and in
the same order (take a look in {\sc IGMtransfer}'s subroutine
\verb+ReadData+ to see how the records are read).

In order to use {\sc IGMtransfer} with AMR, it must be compiled with the flag
``\verb+DAMR+'' (see \sec{comp}).
For a regular, non-adaptive grid, omit the \verb+LevelString+ record and
compile without the \verb+DAMR+ flag. A permitted, but somewhat inefficient,
alternative would be to let \verb+LevelString+ consist solely of
\verb+ni+$\times$\verb+nj+$\times$\verb+nk+ zeros.


\subsubsection{{\tt n\_HIString}}
\label{sec:nh}

Number density $\nhi$ of neutral hydrogen in cm$^{-3}$.


\subsubsection{{\tt TString}}
\label{sec:T}

Temperature of the gas in Kelvin.
Note that in version 1.0 this record contained instead the corresponding
frequency Doppler width.


\subsubsection{{\tt V\_[xyz]String}}
\label{sec:V}

The arrays \verb+V_xString+, \verb+V_yString+, and \verb+V_zString+ contain the
bulk velocity of the gas element in the $x$-, $y$-, and $z$-direction,
respectively, in km s$^{-1}$. The bulk velocity is the sum of the elements'
peculiar motion and the overall Hubble expansion, and are expressed in
physical coordinates with respect to the center of the box. Note that in
version 1.0 this record contained instead the velocities in terms of the
thermal velocity broadening of the Ly$\alpha$ line.


\subsubsection{{\tt ZString} and {\tt n\_HIIString}}
\label{sec:nd}

{\sc IGMtransfer} includes an optional prescription for dust.
The dust cross section is
expressed as a cross section $\sigma_{\mathrm{d}}(\lambda)$ \emph{per hydrogen
nucleus}, with the wavelength
dependence given by a fit to extinction curves of dust in the SMC and the LMC
by \citet{pei92}, \citet{wei01}, and
\citet{gne08}\footnote{{\sc IGMtransfer} uses a faster functional form, seen in
the subroutine {\tt XsecD},
which is accurate to the sub-\permil~scale around the Ly$\alpha$ line, and to
a few percent in the range $[$800,2000$]$ ($[$800,1500$]$) {\AA} for SMC (LMC)
dust.}. Thus, in principle the ``dust density'' $\nd$ should actually
be the hydrogen density $n_{\mathrm{H}}$, such that the optical depth of
dust, as seen by a photon of wavelength $\lambda$ traveling a distance $r$
through a cell of uniform gas and dust density, is
$\tau_{\mathrm{d}} = r \, n_{\mathrm{H}} \, \sigma_{\mathrm{d}}(\lambda)$.
However, since in general the metallicity in
a given cell will differ from that of the reference metallicity $Z_0$
($= Z_{\mathrm{SMC}}$ or $Z_{\mathrm{LMC}}$),
$n_d$ is scaled by the metallicity in that cell, divided by $Z_0$.
Moreover, since arguably dust tends
to be destroyed in regions where hydrogen tends to get ionized, rather than
scaling with $n_{\mathrm{H}}$, in \citet{lau09b} we argued that a more realistic
quantity is obtained by scaling with $\nhi$, plus some fraction of $\nhii$.
Thus, the ``dust density'' in the $i$'th cell is set equal to
\begin{equation}
\label{eq:nd}
n_{\mathrm{d},i} = (n_{\textrm{{\scriptsize H}{\tiny \hspace{.1mm}I}},i}
                 +  f_{\mathrm{ion}}
                    n_{\textrm{{\scriptsize H}{\tiny \hspace{.1mm}II}},i})
                    \frac{Z_i/Z_\odot}{Z_{0,i}/Z_\odot},
\end{equation}
where $Z_\odot$ is Solar metallicity, and 
$0 \le f_{\mathrm{ion}} \le 1$ determines to which
degree the ionized hydrogen fraction contributes to the dust
density: 0 is complete destruction, 1 is complete survival; in
\citet{lau09b} it is argued that setting $f_{\mathrm{ion}} = 0.01$ is a
realistic value, and gives Ly$\alpha$ escape fractions lying roughly midway
between using 0 and 1.
For more details, see
\citet{lau09b}, where the value of $f_{\mathrm{ion}}$ is also thoroughly
discussed.

Thus, to calculate the dust density, two additional records must be provided in
``\verb+CellData+'', namely the metallicity in terms of Solar (\verb+ZString+)
and the density of ionized hydrogen (\verb+n_HIIString+).
Furthermore, the code must be compiled with the flag
``\verb+Ddust+'' (see \sec{comp}).
Note, however, that since by far most of the dust resides in the galaxies,
while the largest part of a given sightline goes through the IGM, in general
the effect of dust is negligible.



\subsection{Galaxy data}
\label{sec:galdata}

Every sightline is assumed to initiate just outside a galaxy, where ``just
outside'' is specified by the keyword \verb+f_rvir+, which is the number of
virial radii from the center.
In \citet{lau11}, we found that the correlation of the velocity field and
density field with the galaxies causes a non-trivial transmission close to the
Ly$\alpha$ line (seen in \fig{spXF}),
and that starting the sightlines $\sim$1.5 virial radii from
a galaxy gives meaningful results. The galaxies in those simulations each
consisted of $\sim$10$^3$ to $\sim$10$^4$ cells.
We emphasize that if your circumgalactic environs are not sufficiently
well-resolved, you will probably not find the same shape of $\Flam$;
the average transmission $\T$, on the other hand, is not so sensitive to
resolution, and does not depend on your chosen value of \verb+f_rvir+.

The normalized spectrum is emitted at rest wavelength in the reference frame of
the
center of mass of a galaxy, which in turn may have a peculiar velocity relative
to the cell at which it is centered. This spectrum is then Lorentz transformed
between the reference frames of the cells encountered along the line of sight.
Since the expansion of space is homologous, each cell can be perceived as lying
in the center of the simulation, and hence this bouncing scheme does not
introduce any bias, apart from reusing the same volume several times for a given
sightline. However, since the sightlines ``scatter'' around stochastically and
thus pierce a given region from various directions, no periodicities arise in
the calculated spectra.

To calculate the above, the appropriate properties of the galaxies are given
in the text file given by the keyword \verb+GalData+, consisting of one line
for each galaxy (plus a ``header'' of an arbitrary number of lines beginning
with ``\verb+#+''), and seven rows of data, explained in \tab{galdata}.
\begin{table}[!t]
\begin{center}
{\sc Contents of ``}{\tt GalData}''\vspace{1mm}
\begin{tabular}{p{2.6cm}p{8.3cm}}
\hline
\hline
Label                & Explanation \\
\hline
\verb+x, y, z+       & $x$-, $y$-, and $z$-position of galaxy in kpc, where the
                       origin is placed in the middle of the box. Note that
                       {\sc IGMtransfer} transforms coordinates from
                       $[-$\verb+D_box+$/2,+$\verb+D_box+$/2]$ to
                       $[0,$\verb+D_box+$]$.\\
\verb+r_vir+         & Virial radius in kpc.\\
\verb+v_x, v_y, v_z+ & $x$-, $y$-, and $z$-components of systemic velocity of
                       galaxy in km s$^{-1}$, in the reference frame of the
                       center of the box. This velocity will be similar to, but
                       in general different from, the bulk velocity of the cell
                       at which the galaxy is centered.\\
\hline
\end{tabular}
\caption{{\small Physical properties of the galaxies at which the sightlines
                 are started. In the example simulation these were identified
                 as bound structures satisfying the following criteria:
                 it must not be a substructure of a larger structure,
                 it must have at least 15 star particles, and
                 it must have circular velocity larger than 35 km s$^{-1}$.
                 These criteria were imposed to ensure sufficient resolution.}}
\label{tab:galdata}
\end{center}
\end{table}

If you are not interested in the properties of the IGM in the vicinity of
galaxies ($\F$), but merely in the average properties of the IGM ($\T$),
a similar file may
be constructed, with random initial positions, \verb+r_vir+ $= 0$, and
\verb+v_x+, \verb+v_y+, and \verb+v_z+ set equal to the distance from the
center of the box times the Hubble constant $H(z)$.


\subsection{Compilation}
\label{sec:comp}

For memory reasons, all parts of the code concerning the AMR structure and the
dust are included as ``preprocessor directives'', and are compiled only if
the appropriate flags are set. To invoke AMR structuring, compile with the flag
``\verb+-DAMR+'' (and include a record in the ``\verb+CellData+'' file with the
refinement levels of the cells); to invoke dust modeling, compile with the flag
``\verb+-Ddust+'' (and include two records with the metallicity and the ionized
hydrogen density, respectively).

Accordingly, to compile {\sc IGMtransfer} for performing RT in the test data
file \verb+testdir/CellData.bin+, which include both AMR and dust, write

\begin{verbatim}
   ifort -O3 -fpp -DAMR -Ddust IGMtransfer.f90 -o IGMtransfer.x
\end{verbatim}
or
\begin{verbatim}
   gfortran -O3 -x f95-cpp-input -DAMR -Ddust IGMtransfer.f90 -o IGMtransfer.x
\end{verbatim}
or
\begin{verbatim}
   g95 -O3 -cpp -DAMR -Ddust IGMtransfer.f90 -o IGMtransfer.x
\end{verbatim}
or
\begin{verbatim}
   pgf90 -O3 -Mpreprocess -DAMR -Ddust IGMtransfer.f90 -o IGMtransfer.x
\end{verbatim}
depending on your compiler. The simulation is then run with
\begin{verbatim}
   ./IGMtransfer.x < test.in
\end{verbatim}

Even if neither of the two optional flags are used, you must still compile
with the flags that enable the preprocessing; that is, to compile with neither
the AMR nor the dust option, use
\begin{verbatim}
   ifort -O3 -fpp IGMtransfer.f90 -o IGMtransfer.x
\end{verbatim}
for the \verb+ifort+ compiler, and similarly for other compilers.



\section{{\sc ProcessIGM}}
\label{sec:process}

The output of {\sc IGMtransfer} is a huge file containing a spectrum for each
sightline, i.e.~a filesize of
4 $\times$ \verb+n_gal+ $\times$ \verb+n_los+ $\times$ \verb+SpecRes+ bytes.
The name of this file is given by the keyword \verb+Ioutfile+ in the
input parameter file.

To extract $\Flam$ and $\T(z)$ from this file, {\sc ProcessIGM} is called with
the same input file as was used for {\sc IGMtransfer}, i.e.
\begin{verbatim}
   ./ProcessIGM.x < test.in
\end{verbatim}
where the compiled executable is assumed to have been named \verb+ProcessIGM.x+.

The transmission function $\Flam$ is determined by calculating in each
wavelength bin the median of all values of $e^{-\tau}$ for all sightlines.
To know the dispersion of the transmission for different sightlines, also the
16th and the 84th percentiles are calculated. The result is written in a
formatted file, the name of which is given by the keyword \verb+Poutfile+ in
the input file, with one
row for each wavelength bin, and the columns giving wavelength, median, 16th,
and 84th percentile. The fifth column is not really used, but gives the average
in each bin.

The average transmission $\T$ is determined by calculating first for each
sightline
$j$ the total transmitted flux in the wavelength interval given by the keyword
\verb+BW_stat+, i.e.
\begin{equation}
\label{eq:T}
\T_j = \frac{1}{n_{\mathrm{bins}}}
                              \sum_{i_1}^{i_2}
                              e^{-\tau(i\mathrm{'th\,bin})},
\end{equation}
where $i_1$ ($i_2$) is the bin number corresponding to the value of
\verb+BW_stat(1)+ (\verb+BW_stat(2)+) and
$n_{\mathrm{bins}} = i_2 - i_1 + 1$ is the number of bins in the applied
interval. Subsequently, the median (and 16 and 84 percentiles) of all
sightlines is calculated. The resulting three numbers are
the three last numbers written to standard output when running
{\sc ProcessIGM}.


\section{{\sc F\_lam}}
\label{sec:F_lam}

The output file from {\sc ProcessIGM}, given by the keyword \verb+Poutfile+ can
be visualized using the IDL code
{\sc F\_lam}, which as input takes the same input file as the two previous
codes:
\begin{verbatim}
   IDL> .r F_lam.pro
   IDL> F_lam, 'test.in' [, outfile=outfile] [, xrange=xrange] [, yrange=yrange]
\end{verbatim}

The three optional arguments are:\vspace{1mm}\\
\begin{tabular}{p{1.1cm}p{10cm}}
\verb+outfile+:& A string giving the name of the figure produced by
                 {\sc F\_lam}. Default value is `\verb+F_lam.eps+'.\\
\verb+xrange+: & A two-element vector giving the range in {\AA}ngstr\"om of the
                 plotted wavelength interval. Default value is [1213,1218].\\
\verb+yrange+: & A two-element vector giving the $y$-range.
                 Default value is [0,1.1].\\
\end{tabular}\vspace{1mm}

If everything has gone well, the output of {\sc F\_lam} should be an
\verb+.eps+ file looking something like the middle panel of \fig{spXF}.


\section{Example simulation}
\label{sec:ex}

The included example files in the subdirectory \verb+testdir+ is a snapshot at
$z \sim 3.5$ from (an improved version
of) the hydro/gravity-simulations described in \citet{som03} and \citet{som06}.
This simulation uses a TreePM/smoothed particle hydrodynamics-technique
rather than AMR. Accordingly, the physical parameters of the particles have
first been interpolated onto an adaptive mesh, using the appropriate smoothing
kernels.
A \citet{haa96}-like UV background has been assumed, but initiating at
$z = 10$ rather than $z = 6$. Furthermore, this simple UV RT scheme has been
supplemented by a more elaborate post-processed ionizing UV RT scheme, as
described in \citet{raz06,raz07}.

\begin{figure}[!t]
\centering
\ifnum\figtype=1 \includegraphics [width=0.90\textwidth] {S29COSMOz2p51.eps} \fi
\ifnum\figtype=2 \includegraphics [width=0.90\textwidth] {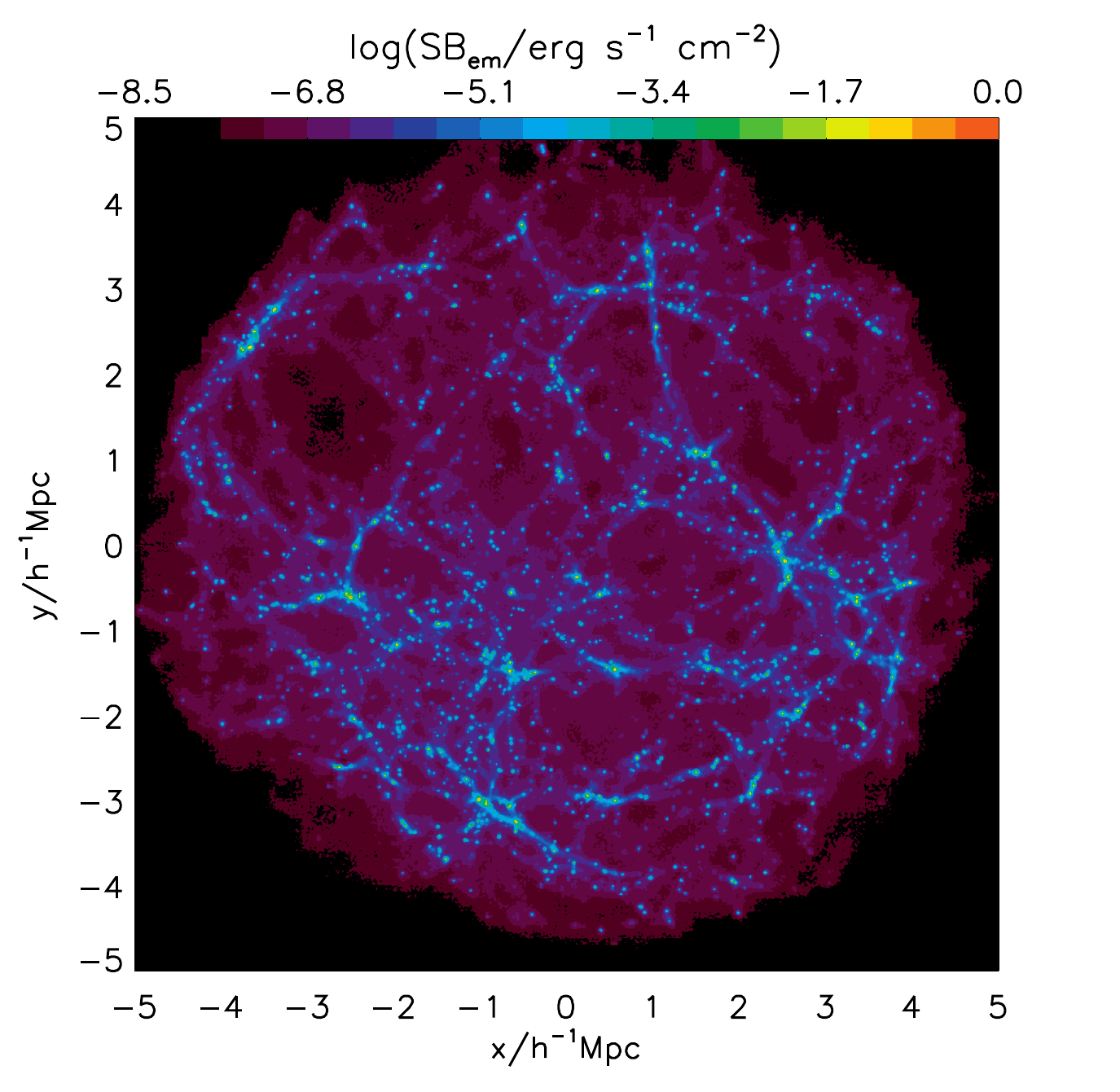} \fi
\caption{{\cap Surface brightness map of Ly$\alpha$ emission in the
               cosmological volume of the example simulation, i.e.~how the
               volume would look if the Ly$\alpha$ radiation escaped directly
               towards the observer.}}
\label{fig:S29COSMOz2p51}
\end{figure}

\Fig{S29COSMOz2p51} shows the simulated volume.

\subsection{Toy model}
\label{sec:toy}

Additionally, a subdirectory \verb+toymodel+ includes an ASCII file with
parameters for the toy grid seen in \fig{AMR_3D}. The purpose of this is
primarily to see how the data are structured; performing the RT in this is
possible, but doesn't really give meaningful results. The code {\sc dat2bin}
converts the ASCII data to binary form in a file that can be read by
{\sc IGMtransfer}. Note that the data do not include dust, so {\sc IGMtransfer}
will have to be compiled with the \verb+DAMR+ flag, but without the
\verb+Ddust+ flag.



\section{Acknowledgments}
\label{sec:ack}

{\sc IGMtransfer} has been developed in collaboration with Jesper
Sommer-Larsen and Alex O. Razoumov. The structure of the hierarchical cell tree
is based on Alex' algorithm for the FTTE RT scheme \citep{raz05}.

The example data files are kindly provided by Jesper, and the
realistic ionizing UV RT was performed by Alex. Thanks.


\section{License and citing}
\label{sec:lic}

{\sc IGMtransfer} and its associated programs are free software, distributed
under the
\href{http://www.gnu.org/copyleft/gpl.html}{GNU General Public License},
i.e.~it may be freely distributed, copied, and even modified, as long as any
changes in distributed versions are indicated.

For applications of {\sc IGMtransfer} leading to publications, please cite
\citet{lau11}.


\end{document}